\documentclass[12pt,preprint]{aastex}
\usepackage{graphicx}
\newcommand{\kms}{km~s$^{-1}$}

\newcommand{\etal}{{\it et al.}}
\slugcomment{European Astronomical Society Publication Series, 2013,\\
  Eds. P. Kervella, Th. Le Bertre \&\ G. Perrin, in press} 
\begin{document}

\title{Direct Ultraviolet Imaging \\
 and Spectroscopy of Betelgeuse}

\author{A. K. Dupree and R. P. Stefanik}

\affil{Harvard-Smithsonian Center for
  Astrophysics,\\ 60 Garden Street, Cambridge, MA 02138 USA}

\begin{abstract}
Direct images of Betelgeuse were obtained over a span of 4 years
with the Faint Object Camera on the Hubble Space Telescope.  These
images
reveal the extended ultraviolet continuum emission ($\sim$2 times the optical
diameter),
the varying overall ultraviolet flux levels and a pattern of bright 
surface
continuum features that change in position and appearance over several months
or less.  Concurrent
photometry and radial velocity measures support the model of a
pulsating star, first discovered in the ultraviolet from
{\it IUE}.   Spatially resolved {\it HST} spectroscopy reveals a larger
extention in chromospheric emissions of Mg II as well as   the rotation
of the supergiant. Changing localized subsonic flows occur
in the low chromosphere that can cover a substantial fraction of the
stellar disk and may initiate the mass outflow. 
\end{abstract}

\section{Introduction}

Alpha Orionis ({\it Betelgeuse}) has been long and well-studied with a variety of ground 
and space-based
techniques as a prototypical supergiant.  Ultraviolet observations have been particularly 
useful because they probe the very outer layers of this star and can pinpoint the onset
of outflowing material and indicate the driving mechanisms behind the mass loss
from the star (Dupree 2010). In fact the monitoring of the flux from {\it Betelgeuse}  shows that
its photometric behavior and its 'spottedness' differ from the signals of magnetic activity
found in the Sun and active cool stars.
The long-lived {\it IUE} satellite clearly demonstrated the presence of
periodic fluctuations and a traveling disturbance in the outer atmosphere 
(Dupree \etal\ 1987).  And the {\it Hubble Space Telescope} with its Faint Object
Camera acquired the first direct image of a star other than the Sun (Gilliland \& Dupree 1996)
which in concert with ground-based photometry and spectroscopy and further ultraviolet
imaging  reveals the extent and characteristics of the supergiant's variability. 

\begin{figure}[!ht]
\vspace*{-0.2in}
\begin{center}
\includegraphics[angle=0,scale=0.6]{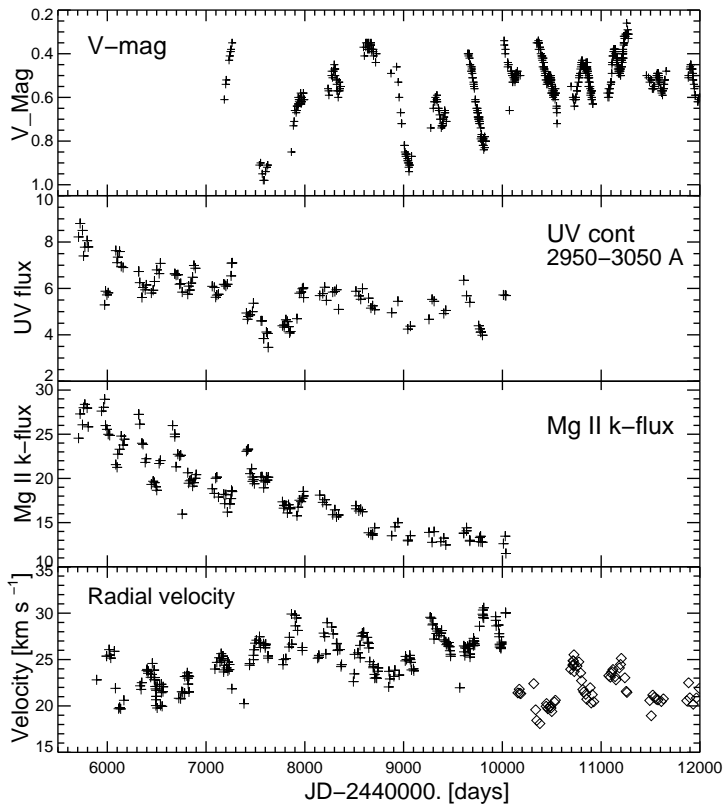}
\vspace*{-1.0in}
\caption{Panels showing the V magnitude from the AAVSO photoelectric
  database (Henden 2012), the UV continuum and the 
Mg II k-line flux  measured from IUE spectra, and the
radial velocity.  In the bottom panel, measurements denoted by the plus-symbol (+) are taken from
Smith \etal\ (1989); the open diamonds ($\diamond$) represent measures
from Oak Ridge Observatory. The radial velocity measures are generally made
from photospheric neutral metal lines. A discontinuity occurs between the two sets of radial velocity
measures; this discontinuity appears to be real, since no cause has been identified.}
\end{center}
\vspace*{-0.3in}
\end{figure}

\section{Spatially Unresolved Photometry and Spectroscopy}

Over a time span of $\sim$ 16 years, photometry in the V-band and the ultraviolet
continuum ($\lambda \lambda$ 2950--3050\AA), the Mg II k-line emission flux, and the
radial velocity are shown in Fig. 1.  For  $\sim$3 years
(JD 6000-7400, mod JD2440000), the chromosphere displayed a brightness fluctuation 
with a period of 420 days (Dupree \etal\ 1987) which later became substantially weaker, and then
disappeared entirely. The appearance of a period suggests that the global brightness variations
do not arise from the appearance of convection cells on the surface which would not be
expected to be periodic. A period of $\sim$400 d is consistent with models
of fundamental pulsation modes of the star (Lovy \etal\ 1984, Stothers 2010).  From  
Fig. 1, it is apparent 
that the  continuum brightenings are correlated with   
the high chromospheric (Mg II) brightenings.  This is also a 
clue that Betelgeuse's behavior is not the result of magnetically active 'star spots'.  Cool
stars with magnetic activity are well-known to show an anti-correlation between
photometric brightenings and chromospheric activity.  When classical star spots
are present, they are cooler and photometrically  'dark', the continuum flux decreases, 
and chromospheric emission
lines become stronger as a result of magnetically-associated chromospheric heating. 
The V-magnitude decreases  during three instances (JD7600, JD9100, JD9800, mod JD2440000) 
where the UV continuum measures, observed  simultaneously with V-band photometry exhibits a 
decreased flux.   Additionally, the flux modulation in Betelgeuse is substantial ... 
about a factor of 2 in the lines and continuum ... and such an excursion surpasses
that found in low gravity magnetically active stars such as RS~Cvn binaries. 

An indication of the presence of a travelling wave in the atmosphere comes from measures
of the B-magnitude variations and the flux in the chromospheric lines 
reported in Dupree \etal\ (1987).  On several occasions during the time when the 420-day period
was evident, the B magnitude became faint and then recovered while, after a delay
of 55 days, the Mg II h-line flux became faint and subsequently recovered.  If a propagating wave 
caused the decrease in emission measure, followed by an increase, and this wave travelled
at $\sim$2 \kms, it would cover a reasonable distance of  0.1 R$_\star$ in 55 days, 
using the larger distance of Betelgeuse (191 pc) suggested by Harper \etal\ (2008).
Observation of the variation of the Mg II h and k lines suggests a lag of $\sim$70 days
beween the h and k line variations.  Because the opacity in the k-line is larger
than the h-line, the k-line is formed further out in the atmosphere and the lag
in flux variation between this two lines again is consistent with a propagating disturbance. 

The radial velocity measures might offer additional information. The values displayed
in Fig. 1 ({\it lower panel}) suggest a long-term variability ($\sim$13 yr) on which
shorter variations ($\sim$400 d) are superposed.    In a Cepheid star,
the light maximum is close to but does not always coincide with the maximum velocity of approach
(cf. Robinson \& Hoffleit 1932; Bersier 2002). One CS Mira star, R CMi,  has shown light maximum
after  maximum
velocity infall (Jorissen 2004, and Lion \etal\ 2013).  
Detailed study shows the velocity pattern in Miras is complex and varies
with the line diagnostic (Hinkle \etal\ 1982).  Inspection of 
Fig. 1 shows an inconsistent  pattern at many light maxima in Betelgeuse.  
For instance, the V-magnitude 
displays brightenings at  JD8600, JD 10800, JD 11200 (mod JD2440000)
coincident with the radial velocity 
corresponding to  a local infall 
maximum - not subsequent to the infall maximum as seen in a Mira variable.  
However, the data at JD9800 exhibit a minimum V-magnitude brightness, yet the radial velocity 
measures signal maximum inflow. Gaps in the observational measures 
obviously can compromise conclusions here.  The spatially resolved observations discussed
later in this paper appear to offer an explanation of the radial
velocity behavior, suggesting that a clean interpretation and seeking 
similarity with other pulsating stars remains challenging.

In sum, the spatially unresolved measures suggest that pulsation phenomena
could dominate the photometric variability of Betelgeuse, but the
details do not consistently 
replicate in detail what is found in globally pulsating Cepheid stars
or in Mira giants.

\section {Direct UV Imaging of Betelgeuse}

The Faint Object Camera on {\it HST} was used (Gilliland \& Dupree 1996) 
to image Betelgeuse directly in 
the ultraviolet continuum ($\lambda$2550\AA).  This first direct image of the
surface of a star other than the Sun provided about 10 resolution elements on the
ultraviolet disk (38 mas point spread function) which has a diameter about 2.2 times
larger than the optical diameter.  This image revealed a bright spot in the SW quadrant
of the star which comprised 10\% of the area and 20\% of the flux from Betelgeuse at that
time. Subsequently, we followed up with similar ultraviolet images spanning 4.1 years. 
All of the images were obtained with a combination of filters: 
a medium-band filter (F253M) was crossed with a second UV filter, F220W, 
and 4 magnitudes of neutral-density filter inserted also. 
These images are shown in Fig. 2  where  different scalings are used
in the upper  and lower panel set.  Each of these
images contains the comparison star, HZ 4, taken during the first visit
of HST ($t = 0$) demonstrating the extended nature of Betelgeuse
in the ultraviolet.

When the images are scaled to the same exposure time (Fig. 2, {\it top 
  panels}), it is obvious that the
total ultraviolet flux not surprisingly varies on a time scale of
  months. The lower panels in Fig. 2  show the
same images scaled to the brightest pixel.  This figure reveals the changing 
brightness pattern across the stellar chromosphere. The single bright area found at $t=0$ 
becomes smaller and fainter over the next 2.6 yr, then appears to move to the
north, and becomes greatly extended in the 3.5 yr observation, approximately 
'circling' the spot in the original ($t=0$) image, before fading in the 
final image at 4.1 yr. The bright spots seem to stay in approximately the same position on the
star.  We find no large excursion to the stellar limb in the position of the bright spot. 

Characteristics of the UV images are shown in Fig. 3 with respect to the V magnitude and the
radial velocity. The mean UV flux from the HST images ({\it third panel}) 
generally tracks the V magnitude.  In fact, 
the faintest excursion in magnitude is in harmony with the lowest value of the UV flux, and
the times of brightest optical magnitude generally agree with the brightest UV flux. 
The relation beween UV flux and radial velocity is not so clear. The highest infall
velocity occurs twice during the HST observations, and at these times
the

\newpage

\begin{figure}[!ht]
\begin{center}
\includegraphics[angle=0, scale=0.5]{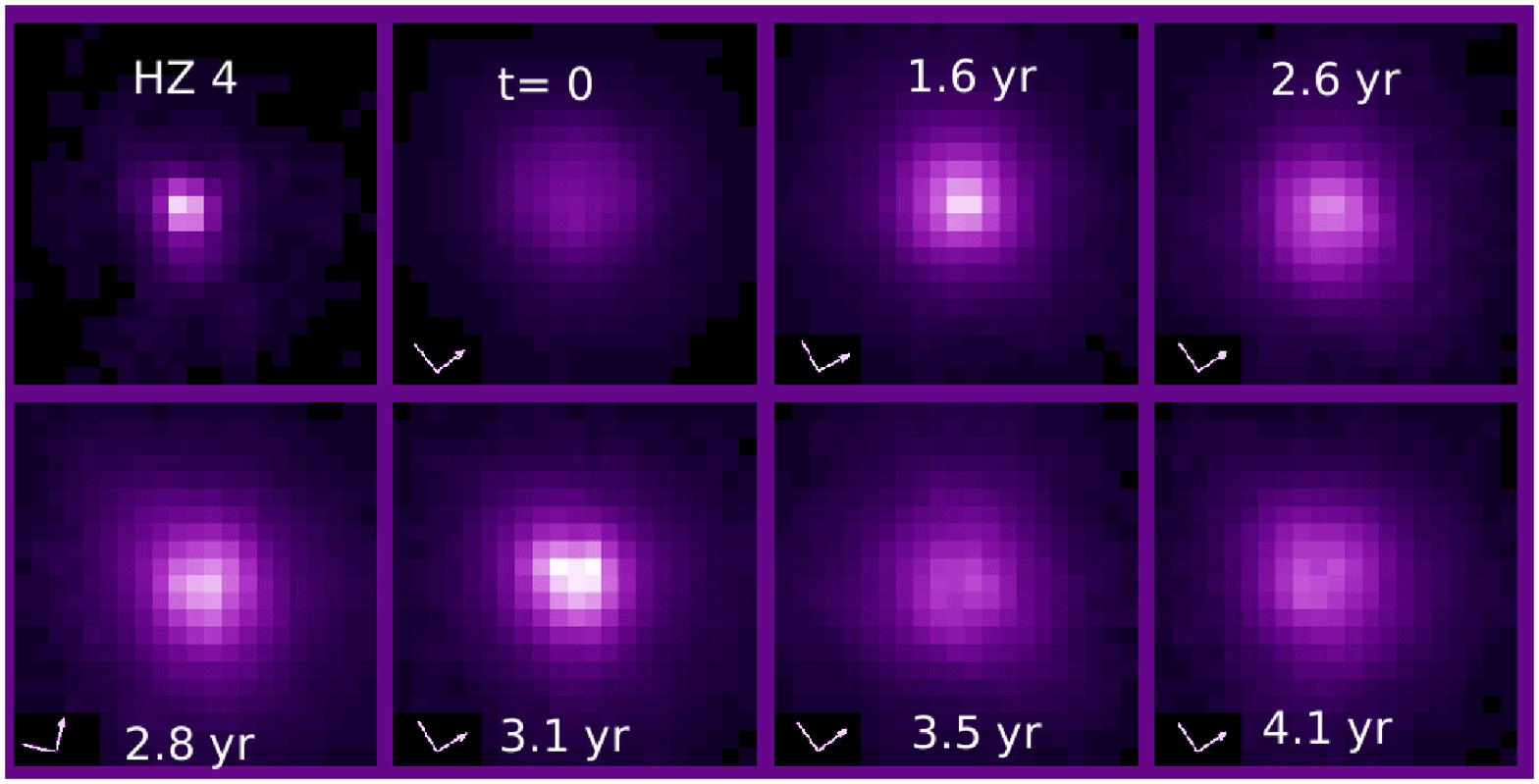}
\includegraphics[angle=0,scale=0.498]{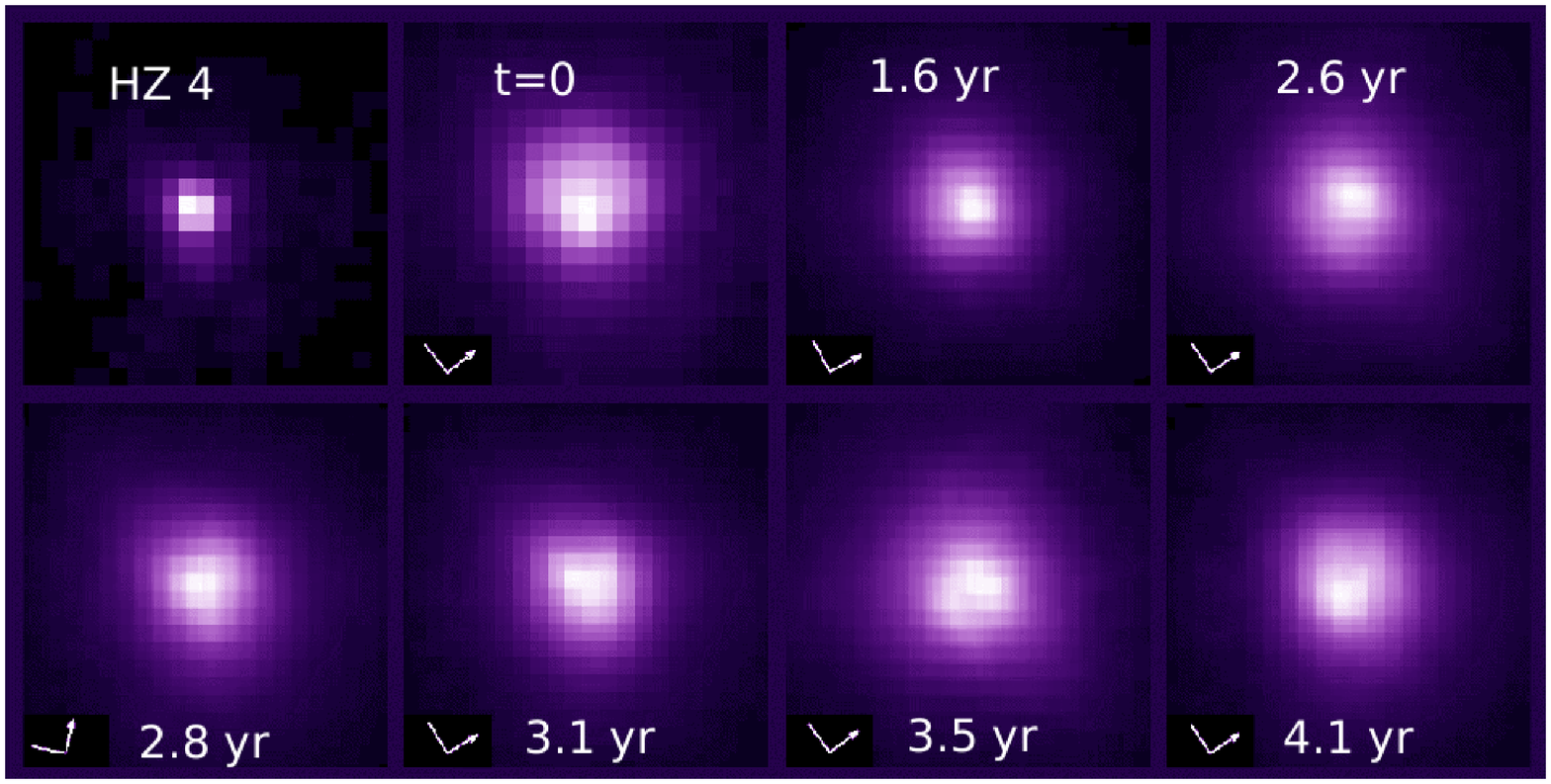}
\caption{{\it Top 2 panels:} UV images  scaled to the same exposure
  time, 3559 s, which corresponds to the longest summed exposure. {\it
  Lower 2 panels:} UV (F253M) dithered
  image scaled to the  brightest pixel.} 
\end{center}
\end{figure}

\noindent
mean UV flux is first at a minimum and then at a maximum value.

A Voigt profile was fit to the UV continuum
images and the full width at half maximum (FWHM) is also shown in Fig. 3 ({\it bottom panel}). 
There does not appear to be a relationship between the diameter of the UV image and
the photospheric radial velocity. During the time span of the HST images, the V magnitude 
displays a period of 366 days whereas the period found for the radial
velocity variation is  440 days. These 
periods were derived using the Lomb-Scargle technique for irregularly-spaced data 
after removing a linear trend
(Horne \& Baliunas 1986).   The absence of correlations may be understood
from the results of the spatially resolved spectra discussed below.

\begin{figure}[!ht]
\begin{center}
\vspace*{-0.5 in}
\includegraphics[angle=0,scale=0.6]{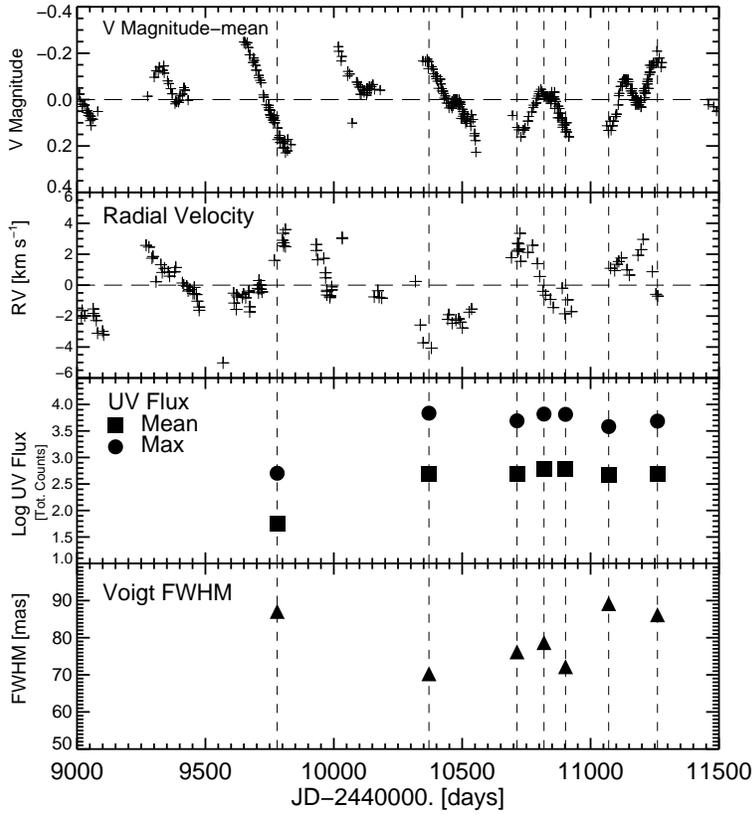}
\end{center}
\vspace*{-1.0 in}
\caption{These panels relate the magnitude and metal-line radial
  velocity
to the HST measures of flux and stellar diameter. The broken lines
are meant to guide the eye.  Long term trends have been removed
with a second-order polynomial from the V magnitude and radial
  velocity
values.}
\end{figure}

\begin{figure}[!h]
\begin{center}
\vspace*{-0.3in}
\includegraphics[angle=0.,scale=0.5]{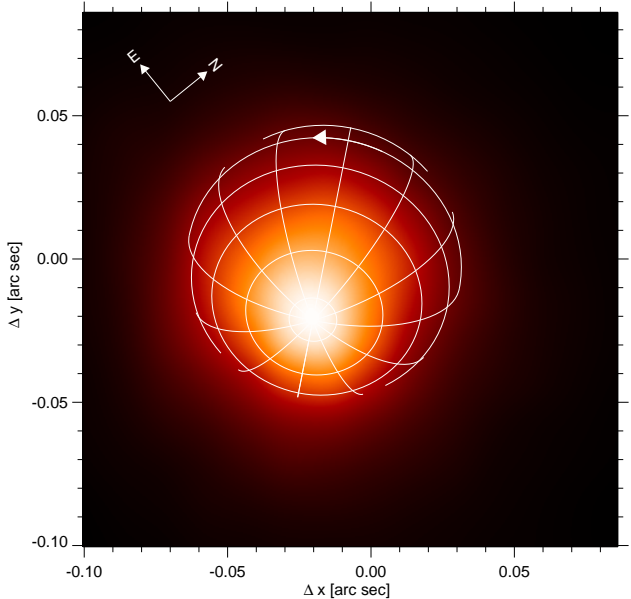}

\caption{UV image of Betelgeuse with the sense of rotation shown as a
wire frame (Uitenbroek \etal\ 1998).}
\end{center}
\end{figure}

At the same time of the first UV image, spatially resolved UV spectra were
obtained (Uitenbroek \etal\ 1998) in the near ultraviolet region that included
both the resonance Mg II emission lines and several photospheric absorption lines.  
The absorption lines from the spatial scan NW to SE across the stellar
disk displayed a systematic shift from negative to positive velocities with
a total amplitude $\sim$10 \kms.  This behavior is interpreted as due to the
rotation of the star. Uitenbroek \etal\ proposed that the bright spot coincided with
the pole of rotation (which is also consistent with the measures of the angle of 
highest polarization), making the inclination of the rotation axis of the star 20$^\circ$ 
from the line of sight (see Fig. 4).  This inclination, coupled with the measured 5 \kms\ rotation velocity
suggests that the deprojected radial velocity is 14.6 \kms, and the rotational period
is 25.5 yr at a distance of 191 pc.  Thus it appears plausible that the bright spots shown in
Fig. 2  emerge preferentially around the pole of the star.

\begin{figure}[!ht]
\begin{center}
\includegraphics[angle=0,scale=0.9]{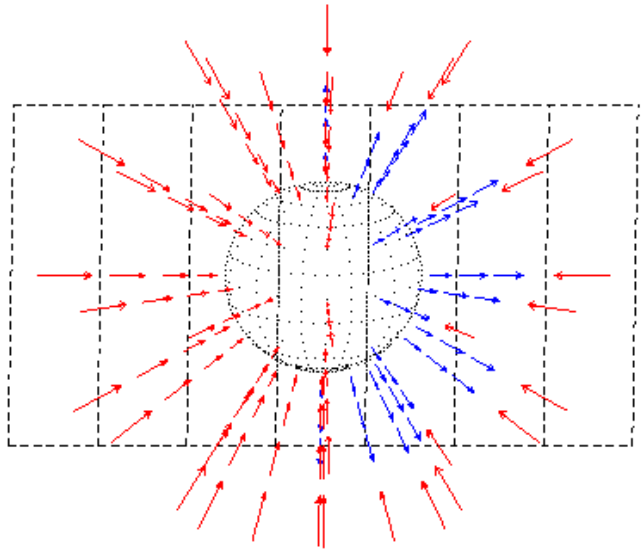}
\end{center}
\caption{The flow velocity in the low chromosphere, as inferred from modeling 
of the Si~I line, $\lambda$2516,
observed with STIS (the 7 STIS aperture positions are  marked by  
broken lines) at time $\Delta t= 4.1 yr$.  The length
and direction of the arrows indicate the magnitude of the velocity and the
direction of flow  (Lobel \& Dupree 2001).}

\end{figure}

\section{Spatially Resolved Spectroscopy}

STIS, the {\it Space Telescope Imaging Spectrograph} on HST possesses  a 
narrow aperture (25 $\times$ 100 mas) which offers true spatial resolution
for the ultraviolet emission lines of Betelgeuse since they have  
a diameter of $\sim$270 mas or larger. In addition, line profiles of
several neutral and singly ionized species that occur in the near-ultraviolet
exhibit centrally-reversed emission which serves as a diagnostic of mass motions
in the atmosphere.  Lobel \& Dupree (2001) detected changes in many profiles with
spatial position on the disk that indicated both outflowing and inflowing 
chromospheric material with velocities $\sim$2 \kms.  These velocities change
with position on the disk and also with time. Detailed non-LTE models in spherical
geometry were
constructed to match the profiles of many lines including Fe I, Fe II, Si I, and Al II.  
The 4 spectroscopic
observations spanned  1.3 yr with sampling of about 0.3 yr. These began at $t=2.8$ yr corresponding
to the times in Fig. 2.    Beginning at $t=2.8$ yr,  
the flow pattern in the low chromosphere  changed from
a global decelerating inflow
to outflow in one quadrant and subsequently the outflow extended to almost the whole stellar
hemisphere (Fig. 5). 
The spatially resolved spectroscopy reveals that the outer atmosphere of Betelgeuse
does not behave in a global fashion but that asymmetric time dependent dynamics
are present.

\section{Conclusions}
When observed as a star, the photosphere and chromosphere of Betelgeuse are subject to 
a semi-periodic travelling oscillation with a period of $\sim$400 days.  This
period can be coherent for $\sim$4 years. 

Spatially resolved imaging shows that the image size in the 
near ultraviolet continuum ($\lambda$ 2550\AA) exceeds the optical
diameter (taken as 55 mas) by about a factor of 2.2, and the
chromospheric 
Mg II lines extend even further ... to
a diameter $\sim$4 times that of the optical. Bright regions occur on the ultraviolet 
disk that change in position and strength over a period of
months.  They appear to be localized around the rotational pole of the star. 
Spatially resolved spectroscopy demonstrates that the low chromosphere
does not behave uniformly, but that the dynamics are complex. We have discovered 
gradually changing inflow and outflow patterns suggesting asymmetric mass motions.
Such behavior obviously complicates the interpretation of the spatially unresolved radial
velocity measures.  
 
We acknowledge with thanks the variable star observations from the
AAVSO International Database contributed by observers worldwide and
used in this research.


\end{document}